# tinyLTE: Lightweight, Ad Hoc Deployable Cellular Network for Vehicular Communication

Fabian Eckermann, Philipp Gorczak and Christian Wietfeld

TU Dortmund University, Communication Networks Institute (CNI)

Otto-Hahn-Str. 6, 44227 Dortmund, Germany

E-Mail:{fabian.eckermann, philipp.gorczak, christian.wietfeld}@tu-dortmund.de

*Abstract*—The application of LTE technology has evolved from infrastructure-based deployments in licensed bands to new use cases covering ad hoc, device-to-device communications and unlicensed band operation. Vehicular communication is an emerging field of particular interest for LTE, covering in our understanding both automotive (cars) as well as unmanned aerial vehicles. Existing commercial equipment is designed for infrastructure making it unsuitable for vehicular applications requiring low weight and unlicensed band support (e.g. 5.9 GHz ITS-band). In this work, we present tinyLTE, a system design which provides fully autonomous, multi-purpose and ultra-compact LTE cells by utilizing existing open source eNB and EPC implementations. Due to its small form factor and low weight, the tinyLTE system enables mobile deployment on board of cars and drones as well as smooth integration with existing roadside infrastructure. Additionally, the standalone design allows for systems to be chained in a multi-hop configuration. The paper describes the lean and low-cost design concept and implementation followed by a performance evaluation for single and two-hop configurations at 5.9 GHz. The results from both lab and field experiments validate the feasibility of the tinyLTE approach and demonstrate its potential to even support real-time vehicular applications (e.g. with a lowest average end-to-end latency of around 7 ms in the lab experiment).

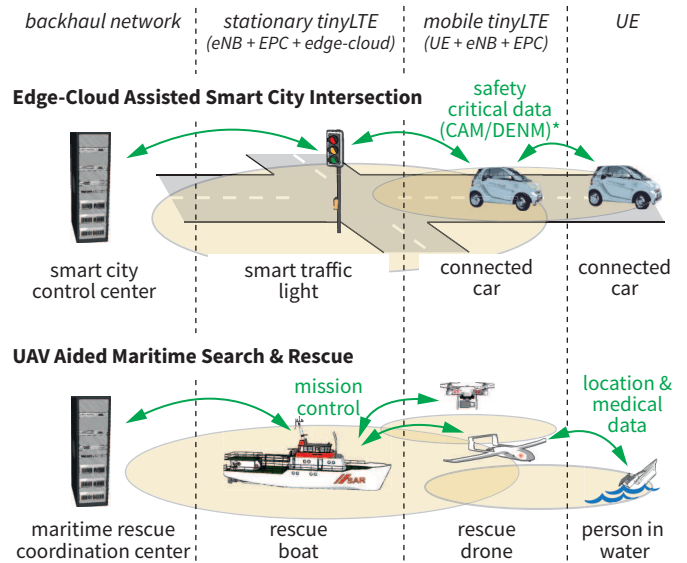

Fig. 1. tinyLTE use cases: a maritime search and rescue and an urban smart intersection scenario. (* CAM: Cooperative Awareness Message, DENM: Decentralized Environmental Notification Message)

## I. INTRODUCTION

Vehicular communication systems need to be deployable in a large variety of different environments ranging from rural areas to dense city centers to maritime settings. At the same time, the great variety of application areas ranging from civil aviation to urban car traffic entails strong requirements [1] [2]. Quality of service (QoS) has to be guaranteed e.g. to ensure navigational collision avoidance and systems need to be highly scalable in terms of communication range and number of communicating nodes. Current generation cellular technologies like Long Term Evolution (LTE) have been designed to provide the aforementioned technical features. However, the static and centralized core network (Evolved Packet Core - EPC), that is responsible for management and authentication tasks of cellular networks, is a source of inflexibility. Indeed, LTE-type centralized processing incurs an uplink delay that prevents application to demanding vehicular use cases [3]. Furthermore, coverage depends on the availability of (fixed) base stations. A network partition between a base station and the EPC would render the base station unable to provide any service to user equipment. Recent studies analyzed the feasibility of current LTE networks for unmanned aerial vehicles (UAV) and state that mobility enhancements are necessary [4].

Unlicensed, distributed technologies such as IEEE 802.11 offer great deployment flexibility via mesh or ad hoc modes. However, they work in unlicensed bands with uncontrollable interference and are limited with respect to security and quality of service [5] [6].

A lightweight, low cost LTE network with ad hoc capabilities is a promising technology for vehicular communication. Possible use cases include drone- or car-traffic and especially emergency scenarios like disaster recovery and search & rescue as exemplified in Fig. 1. Cellular networks' service guarantees could be kept while adding some of the flexibility of mesh networks. Furthermore, a reduced dependency on intermediate communication with a centralized entity adds the potential of delivering very low average latency. Previous studies have found a fully centralized concept to limit robustness in disaster scenarios, motivating the implementation of eNB for use as aerial base stations with partial autonomy [7]. Some autonomous LTE base stations are commercially available but only operate in a limited number of frequency bands and their form factors still prohibit integration into UAV and sometimes even into cars.

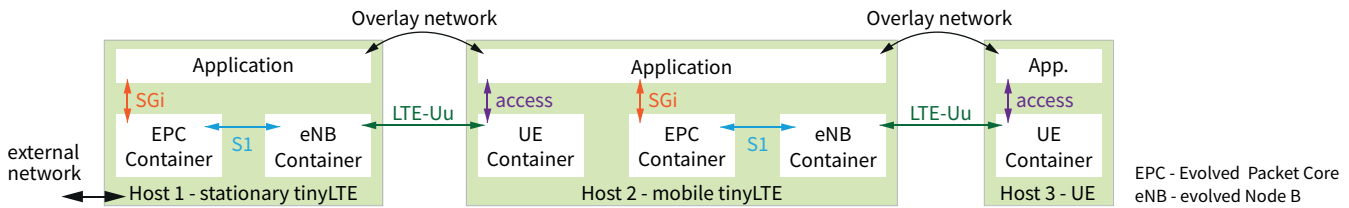

Fig. 2. System design of a two-hop LTE communication by tinyLTE.

Relaying operation has been standardized in release 10 of the LTE specification [8] while UAVs acting as relay stations have been extensively covered in [7] [9]. Although pre release 10 user equipment (UE) is compatible with relay nodes (RN), the standard only supports two-hop operation (via a single RN) and does not consider mobile RN [10]. Standard LTE relaying does not break centralization: While the RN appears just like a base station (Evolved Node B - eNB) to UE, its interface to the core network (S1) is proxied by the base station to which the RN itself is attached (Donor eNB - DeNB). The RN acts as a non-transparent extension to its (fixed) DeNB. The scope of these standardized relaying concepts is motivated by the goal of using RNs as alternatives to conventional eNBs featuring low site acquisition costs and a wireless backhaul [11].

Device-to-device (D2D) communication has been standardized for cellular networks as well starting with LTE release 12. Such direct communication can be divided into inband and outband modes. The inband mode uses licensed LTE bands and can be implemented in two different ways: either by sharing radio resources between cellular and direct communication (underlay), or by allocating dedicated resources for cellular and D2D communication (overlay) [12]. In outband mode, the D2D links utilize unlicensed spectrum and are either controlled by the cellular network or use a random access scheme. Unlike our approach, inband D2D and controlled outband D2D rely on a (static) eNB to control the radio resources. For the distributed access of outband D2D communication on the other hand it is hard to achieve a global optimum [13].

With tinyLTE we combine LTE eNB, EPC and UE on a single device. Each such device is an UE but can also act as a small, fully autonomous cell. Due to this design choice, adding hops does not increase complexity as in the existing LTE-relaying approaches introduced above. We use an IP-based overlay network to implement device-to-device communication and external network access at the application layer. Our integrated approach on a single device yields a major decrease in latency. However, the system operation is CPU intensive and its reliability depends mainly on that of the LTE software stack.

The proposed system uses open source software and runs on standard PC hardware and Linux-based operating systems. Unlike commercial implementations tinyLTE uses software defined radios (SDR) as RF frontends, thus opening up a wide range of possible bands to use. In addition, our solution can be deployed as an LTE relay node in outband mode enabling further applications such as coverage extension of third-party cells. Through outband operation, interference with the root cell is minimized and we can connect vehicles that communicate in any band, for example in the unlicensed spectrum (ISM), to a network operating on standard LTE bands[1].

In the remainder of this paper, we build upon this motivation and explain the underlying system design of tinyLTE (Section II). Section III contains the experimental results of our V2X setup. We finally conclude the paper in Section IV.

## II. SYSTEM DESIGN

We propose a multi-tiered network using frequency multiplexed, outband type 1 device-relaying [14]. By making nodes fully autonomous, we are able to implement a decentralized, IP-based overlay network on top of the cellular communication layer. Each node in this network plays multiple roles: it may act as LTE-infrastructure and LTE-client, while running applications that communicate via the overlay network. In the following section, we first introduce the internal design of the nodes and then describe the three levels of abstraction leading to the overlay network implementation.

### A. Node Implementation

As described earlier, tinyLTE nodes can act as LTE clients and infrastructure. In order to enable flexible provisioning and configuration, we run each key software component in its own virtual environment as indicated in Fig. 2. The UE and eNB components use slightly modified versions of the software stacks from the srsLTE project [15]. Both can interface with a variety of SDR as radio frontends. The EPC component is implemented by a minimal open source implementation[2] which we adjusted for our use case.

### B. Network Layers

As visualized in Fig. 2, each device includes two internal virtual network segments through which containers are able to communicate. The *S1* segment connects the EPC and eNB containers while the *SGi* segment connects the EPC container to the host operating system (and applications).

---

[1]We maintain tinyLTE under https://github.com/tudo-cni/tinyLTE
[2]https://github.com/mitshell/corenet

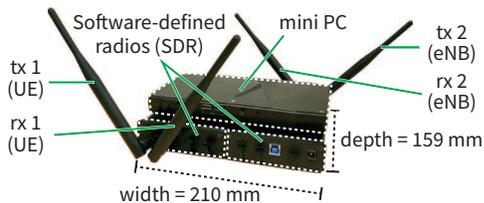

Fig. 3. Hardware in mobile configuration (two SDRs) for experimental setup (width 210 mm, depth 159 mm, height 58 mm; weight 1.4 kg)

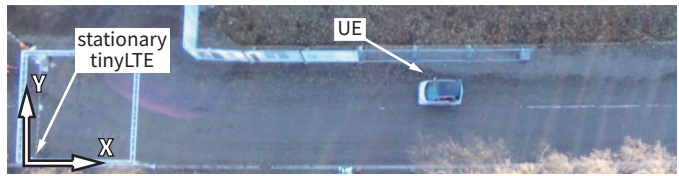

Fig. 4. Aerial view of our measurements for the range extension use case in the field test.

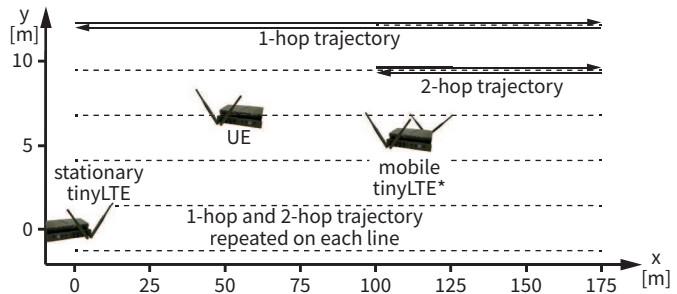

Fig. 5. Trajectories of our experimental setup for a range extension use case. (* only for two-hop trajectory)

On the cellular level, a node may be part of two wireless networks. In that case, it is both a UE in a parent network and the eNB/EPC of the small cell it provides. The external air interface *LTE-Uu* connects UE-containers to eNB-containers across nodes. This results in a hierarchical topology where each node is a client in a parent network and provides connectivity to zero or more child clients. At this level—as in regular LTE operation—the EPC opens a tunnel to the UE using the GPRS tunneling protocol (GTP). In our system, one end of this tunnel is within the EPC container. On the other side, the UE container is privileged to create the corresponding network interface (*access*) directly on the host.

As a final level of abstraction, IP-based routing provides a common overlay network at the application layer. tinyLTE nodes have fixed IP addresses in the overlay network which are assigned through a fixed MMSI-IP-address mapping. Communication is implemented through generic routing encapsulation (GRE) tunnels between hosts. These tunnels terminate at the SGi-interface on the eNB-host, and at the access-interface on the UE-host. At the eNB side, the EPC-container is configured to forward incoming SGi-traffic to its end of the GTP-tunnel described above.

## III. EXPERIMENTAL RESULTS

Example hardware in a mobile configuration is shown in Fig. 3. For laboratory system performance measurements we evaluated a wired setup with both direct communication and two-hop communication. To analyze the system performance in a real-world scenario, we performed experiments motivated by a V2X range-extension use case: a mobile tinyLTE node extends the range of a infrastructure base station (stationary), both nodes working in the 5.9 GHz band. The bandwidth of the LTE cells was set to 5 MHz. We chose an outband relay mode to avoid interference between the stationary and the mobile node. The frequency allocation is summarized in TABLE I.

At first we measured the coverage and the reference signal received power (RSRP) of the stationary node. We then placed a static node so that the signal quality is just good enough to allow a reliable connection to the stationary node. We finally measured the range and the signal quality of the stationary node starting from the mobile node. In each case, end-to-end delays between hosts are measured on the application layer i.e. via the overlay network. We take half of the round-trip time between the communication nodes—measured via ICMP-messages—to determine the mean one-way latency of the system.

### A. Laboratory Setup

For the static laboratory setup we connected the RF frontends of tinyLTE through HF cables and 30 dB attenuators. We then measured the latency and the throughput of tinyLTE for a single-hop and a two-hop connection.

### B. Field Experiments

Our experimental setup was located on a street section closed to the public (Fig. 4). The stationary tinyLTE node was fixed at a height of ∼2 m. Due to the limited power output of our SDR (USRP B210) and the low height, the cell radius of our stationary node was approximately 175 m. Our mobile node was attached to the rooftop of a car (∼1.65 m) resulting in a smaller cell radius. As shown by Fig. 5 we divided the road into six lanes and followed each lane in both directions with a constant speed of ∼1.4 m/s (5 km/h). After the finishing the single-hop measurements, a mobile node is placed at a distance of ∼110 m to the stationary node. We then repeated the drive tests for this two-hop setup starting at the mobile node.

### C. Results

We observed that the achievable throughput highly depends on the performance of the virtualized software stacks. However the low data rate offers some space for optimization of the underlaying software. Adding a second hop did not affect the average throughput.

TABLE I
LTE FREQUENCY ALLOCATION USED IN EXPERIMENTS

| tinyLTE Node Type | Uplink [MHz] | Downlink [MHz] | Duplex Mode |
|---|---|---|---|
| stationary | 5855–5860 | 5895–5900 | FDD |
| mobile | 5865–5870 | 5905–5910 | FDD |

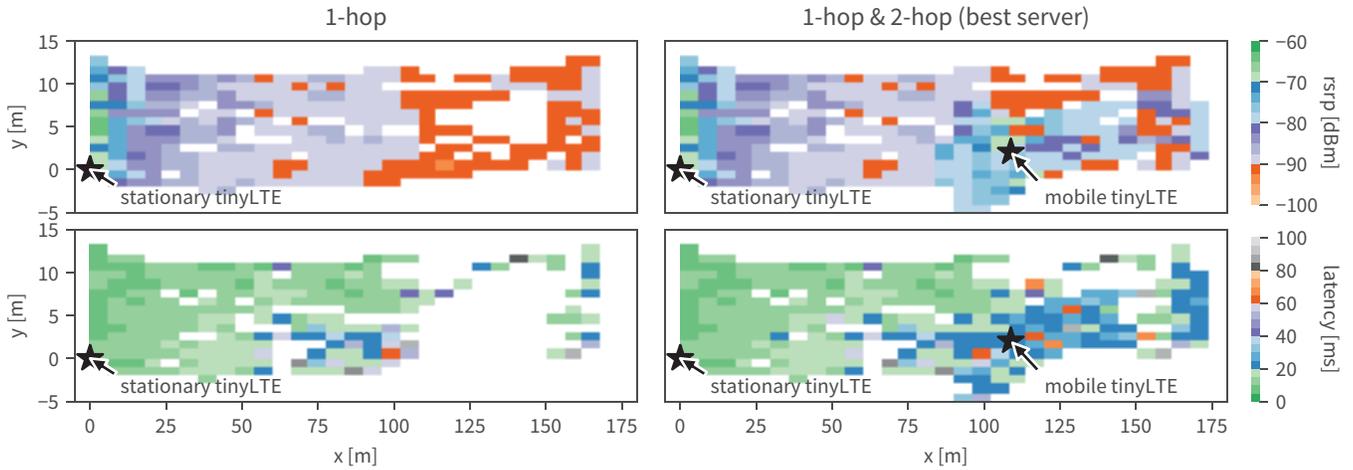

Fig. 6. RSRP and latency evaluation for single-hop measurements and best server plot for single-hop and two-hop. For each tile ($4\,\text{m} \times 2\,\text{m}$) the median of the measured data is calculated.

By tracking the GPS position of the cars, we create a statistical signal strength and latency map of our measurements (Fig. 6). For the single-hop scenario a continuous degradation of the signal strength can be observed as the distance to the stationary node increases. At 110 m the connection becomes very unreliable, while at 160 m the signal strength rises again. The comparison with a two-ray ground model as shown by Fig. 7 suggests that the weak signal strength at a distance of 110-160 m is caused by interference of a strong ground reflection. Similar characteristics have been observed in previous studies of vehicular communications at 5.9 GHz and 5.2 GHz [16] [17]. In the best server plot we compare the median performance measures of both single-hop and two-hop experiments and select the better one for each tile. With the mobile tinyLTE node placed at the edge of the dead spot we can extend the coverage by approximately 50 m as shown by the RSRP best server plot.

The mobile cell improves coverage by filling the aforementioned dead spot. In the best server scenario, the UE is able to communicate with the stationary node across nearly the entire experimental area. The latency map shows that the higher coverage comes at the cost of approximately double latencies due to the additional communication hop. Note that all maps show improved service towards the far edge of the road section. This can be partially explained by a zone of constructive interference of the ground reflection but also by additional time spent in that area during turns (see receive power and density of data points in Fig. 7 and trajectory layout in Fig. 5).

A comparison of the latency in laboratory measurements, the real-world field test and a public LTE network is shown in Fig. 8. In the laboratory setup we achieved an average latency of 7 ms for a single-hop and 13.55 ms for a two-hop communication. In the field tests these values are almost doubled (12.75 ms for single-hop and 22.55 ms for two-hop) due to non-ideal radio channel conditions. Our latency measurements in a public LTE radio network yielded an average of 37.5 ms. It is worth noting that the one-hop latency of tinyLTE consistently outperforms the public LTE network, and even the two-hop setup achieves faster responses with a 90 % probability.

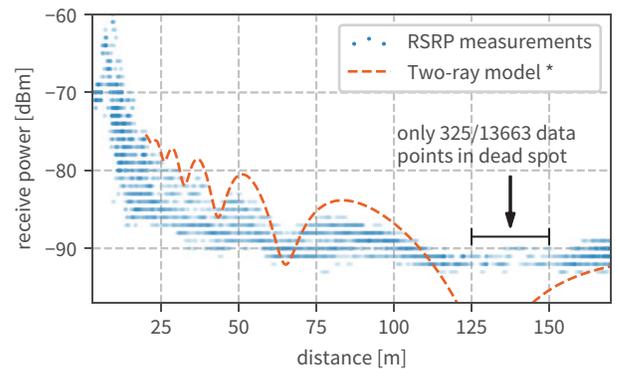

Fig. 7. RSRP measurements collected during the 1-hop experiment show characteristics of a two-ray ground reflection model (* model with wet ground parameters ($\sigma = 2 \times 10^{-2}$ S, $\epsilon_r = 30$) [18])

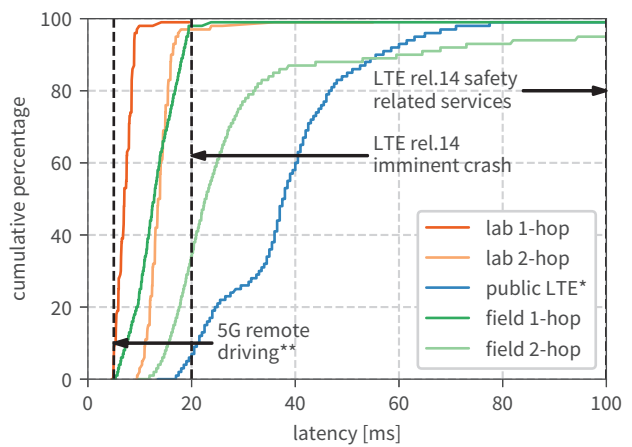

Fig. 8. Cumulative relative frequency plot of end-to-end latency for a wired laboratory setup (with 30 dB attenuator), a public LTE network and a field test. (* Own measurements in public LTE network; ** 5G remote driving requirements see [19])

The 3GPP V2X standard states a maximum latency of 100 ms for safety-related V2X communication and 20 ms for imminent crashes with different message reception reliabilities of 80-95 % [20]. Recent studies [3] [21] compare these requirements to LTE based V2X-implementations, based on prior simulative performance evaluations [22]. It is worth noting that our experimental results are comparable or better than these simulative results. Except for the two-hop field test measurement even the 20 ms V2X imminent crash service requirements are achieved. Although we highly improved the latency within our integrated approach the 5G requirements can not be fulfilled due to essential differences in the underlaying system design.

Our results indicate that co-locating an autonomous core network with the radio access equipment can be a viable design decision for low latency applications. In the single-hop scenario, the near-instantaneous communication between eNB, EPC and edge-cloud applications—all running on the stationary node—yields major improvements in end-to-end latency when compared to communication through a public network with applications running on a remote server.

## IV. Conclusion

In this paper we introduced and evaluated tinyLTE, a low cost, lightweight autonomous LTE network. Building upon virtualization techniques and open source software stacks, our design is a pragmatic approach for building standalone LTE cells with off the shelf hardware. To benchmark our concept we measured system performance in a laboratory setup and real-world field tests. Our experimental results show that the system is able to satisfy the low latency and flexibility demands of vehicular communication. Compared to measurements from a public LTE network tinyLTE exceeds the former one regarding end-to-end latency. Further work is underway regarding the reliability of the two-hop connection and the throughput for both single and two-hop. On the subject of highly dynamic, mobile nodes, we also plan to investigate and improve the impact of frequent connection establishment on end-to-end latency.

## Acknowledgment

The work on this paper has been partially funded by the federal state of Northrhine-Westphalia and the European Regional Development Fund (EFRE) 2014-2020 in the course of the InVerSiV and CPS.HUB/NRW projects under grant numbers EFRE-0800422 and EFRE-0400008, by the German Federal Ministry of Education and Research (BMBF) in the project LARUS (13N14133), by Deutsche Forschungsgemeinschaft (DFG) within the Collaborative Research Center SFB 876 project B4 and the AutoMat project, which received funding from the European Unions Horizon 2020 (H2020) research and innovation program under the Grant Agreement No 644657.